\documentclass[final]{cvpr}

\usepackage{times}
\usepackage{epsfig}
\usepackage{graphicx}
\usepackage{amsmath}
\usepackage{amssymb}

\usepackage{booktabs}
\usepackage{makecell}
\usepackage{multirow}
\usepackage{tablefootnote}

\usepackage[numbers]{natbib}

\usepackage[pagebackref=true,breaklinks=true,colorlinks,bookmarks=false]{hyperref}
\usepackage[flushleft]{threeparttable} 

\newcommand{\Bmat}{{\bf B}}
\newcommand{\Cmat}{{\bf C}}
\newcommand{\Dmat}{{\bf D}}

\newcommand{\Gmat}[0]{{{\bf G}}}
\newcommand{\Hmat}[0]{{{\bf H}}}

\newcommand{\Xmat}{{\bf X}}
\newcommand{\Ymat}[0]{{{\bf Y}}}

\newcommand{\gv}[0]{{\boldsymbol{g}}}
\newcommand{\hv}[0]{{\boldsymbol{h}}}

\newcommand{\xv}{\boldsymbol{x}}
\newcommand{\yv}{\boldsymbol{y}}

\newcommand{\Phimat}{\boldsymbol{\Phi}}

\newcommand{\tsp}{^{\top}}

\def\cvprPaperID{8381} 
\def\confYear{CVPR 2021}

\newcommand\blfootnote[1]{%
	\begingroup
	\renewcommand\thefootnote{}\footnote{#1}%
	\addtocounter{footnote}{-1}%
	\endgroup
}

\begin{document}

\title{Memory-Efficient Network for Large-scale Video Compressive Sensing}

\author{Ziheng Cheng, Bo Chen*, Guanliang Liu, Hao Zhang, Ruiying Lu and Zhengjue Wang\\
Xidian University\\
{\tt\small zhcheng@stu.xidian.edu.cn, bchen@xidian.edu.cn}\\ {\tt\small \{lgl\_xidian,zhanghao\_xidian,ruiyinglu\_xidian,zhengjuewang\}@163.com}
\and
Xin Yuan*\\
Bell Labs\\
{\tt\small xyuan@bell-labs.com}
}


\maketitle

\blfootnote{* Corresponding authors.
}
\begin{abstract}
Video snapshot compressive imaging (SCI) captures a sequence of video frames in a single shot using a 2D detector.
The underlying principle is that during one exposure time, different masks are imposed on the high-speed scene to form a compressed measurement. With the knowledge of masks, optimization algorithms or deep learning methods are employed to reconstruct the desired high-speed video frames from this snapshot measurement.
Unfortunately, though these methods can achieve decent results, the long running time of optimization algorithms or huge training memory occupation of deep networks still preclude them in practical applications.
In this paper, we develop a memory-efficient network for large-scale video SCI based on {\bf \em multi-group reversible 3D} convolutional neural networks.
In addition to the basic model for the grayscale SCI system, we take one step further to combine demosaicing and SCI reconstruction to directly recover color video from Bayer measurements.
Extensive results on both simulation and real data captured by SCI cameras demonstrate that our proposed model outperforms previous state-of-the-art with less memory and thus can be used in large-scale problems.
{The code is at \url{https://github.com/BoChenGroup/RevSCI-net}.}
\end{abstract}

\section{Introduction}

\begin{figure}[!t]
		\centering
        \includegraphics[width=1.0\columnwidth]{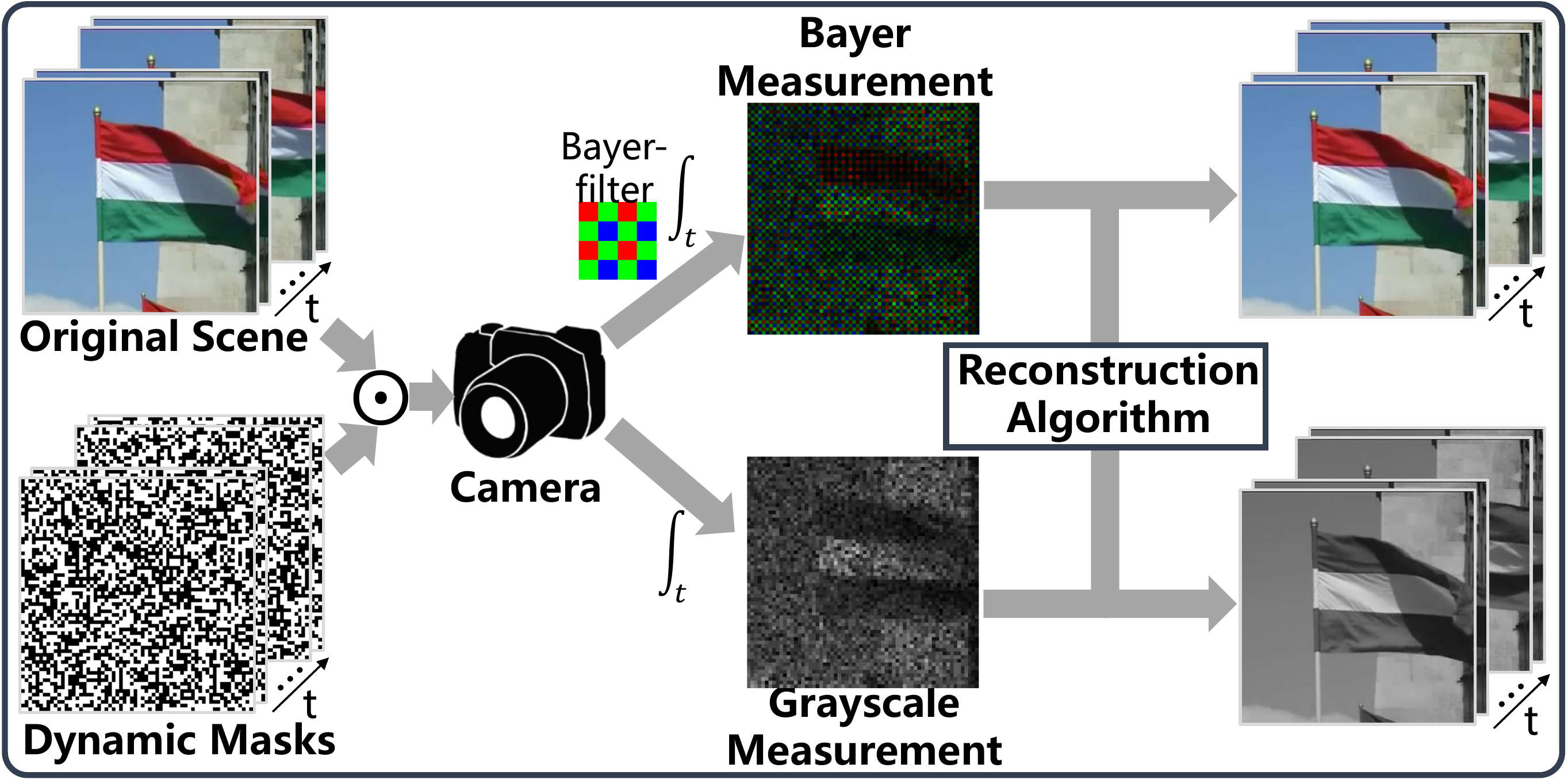}
		\caption{\label{fig:system}Principle of grayscale or color video SCI system. The original scene is modulated by dynamic masks and then is integrated by the camera to obtain a snapshot measurement. Note that for the color video SCI system, the camera captures the modulated scene through the Bayer-filter not directly collecting the brightness. Having obtained measurements, the reconstruction algorithm recovers the desired video from it.}
\end{figure}

Computational imaging (CI)~\cite{Altmann18Science,Mait18CI} introduces modulation (coding) in the optical path to advance the capability of traditional cameras.
Snapshot compressive imaging (SCI)~\cite{Hitomi11ICCV,Patrick13OE,reddy2011p2c2,Wagadarikar09CASSI,Yuan2021_SPM} is a promising CI technique that indirectly captures 3-dimensional (3D) data using a 2D detector, \ie, the original 3-dimensional data (videos or hyperspectral images) are coded by different masks and then integrated into a single frame (measurement).
As shown in Fig.~\ref{fig:system}, in video SCI, the temporal dimension is modulated and compressed, which avoids large memory storage and transmission bandwidth during imaging. 
To make the SCI system practical, an efficient reconstruction algorithm, \ie, recovering the desired images from the compressed measurement is critical.
In this work, we focus on the practical video SCI reconstruction algorithm that can scale to large data.

\begin{figure}[tbp!]
		\centering
        \includegraphics[width=1\linewidth]{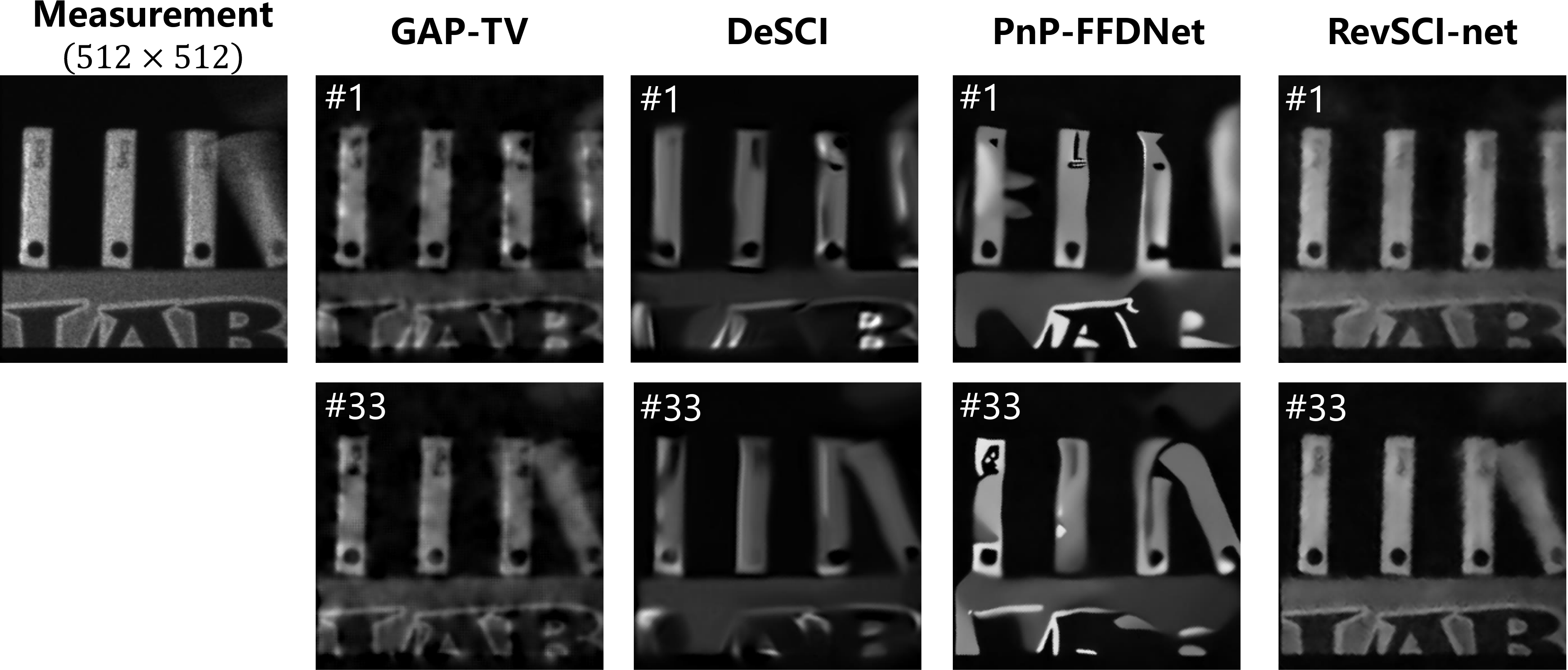}
		\caption{\label{fig:real results} Reconstruction results using the proposed RevSCI-net on large scale (512$\times$512$\times$50) real data captured by~\cite{Qiao2020_APLP}. Note that this is the first deep learning model can perform a compression rate at $B$=50. Videos are shown in the supplementary material (SM).}
\end{figure}

The mainstream of reconstruction methods is the model-based optimization problems with various prior knowledge, \eg, total variation (TV) used in GAP-TV~\cite{Yuan16ICIP_GAP} and TwIST~\cite{ Bioucas-Dias2007TwIST}, and non-local low-rank~\cite{Gu14CVPR} used in DeSCI~\cite{Liu18TPAMI}. These methods can provide usable results in an unsupervised manner but cannot balance the reconstruction quality and speed (hours for DeSCI to reconstruct a $256 \times 256 \times 8$ video from a single measurement), which makes them unrealistic for real applications.
Inspired by deep learning and rich datasets, some researchers develop deep neural networks (DNNs)~\cite{Cheng20ECCV_BIRNAT,Iliadis18DSPvideoCS,Miao19ICCV,Qiao2020_APLP,2016arXivVideoCS} or combine DNN with optimization methods, such as deep unfolding technique~\cite{Li2020ICCP,Ma19ICCV,Meng_GAPnet_arxiv2020} and plug-and-play algorithms~\cite{Yuan20PnPSCI,PnP_SCI_arxiv2021}, to reconstruct desired 3D data from SCI measurements. Benefiting from the efficient feed forward networks, DNN based algorithms decrease the inference time significantly to less than one second.
Most recently, BIRNAT~\cite{Cheng20ECCV_BIRNAT}, which develops a bidirectional recurrent neural network, has led to state-of-the-art reconstruction results.
Most researches on video SCI reconstruction usually verify the performance on a low-resolution situation (less than $512\times512$). With the high-resolution images widely used in our daily life (HD with $1280\times720$ and FHD with $1920\times1080$), however, the squarely increasing pixels will significantly increase the running time and training memory consumption for DNNs.
Although some methods can provide superior reconstruction, \eg, DeSCI and BIRNAT, unpractical running time or GPU memory consumption still preclude them in the practical large-scale SCI system applications.

Bearing the above concerns in mind, in this paper, we propose an {\em end-to-end reversible 3D} convolutional neural network (CNN) for SCI reconstruction named RevSCI-net, in which 3D convolutional kernel jointly explores the {\em spatial and temporal correlation} within the desired data. Meanwhile, the reversible structure allows activations not to be stored in memory. The main contributions of our work are summarized as follows:
\begin{itemize}
    \item Since the desired signal of video SCI is 3D, we build an end-to-end 3D CNN paradigm for video SCI reconstruction which jointly explores the spatial and temporal correlation of video frames by the 3D convolutional kernel. To our best knowledge, this is the first time that 3D CNN is applied in SCI problems.
    \item We propose multi-channel reversible CNN in the proposed network with less memory occupation during training.
    {Benefit by this, we can reconstruct a $512\times512\times50$ video from a snapshot measurement, with an example shown in Fig.~\ref{fig:real results}, where a compression rate of 50 is achieved.} This is the first deep learning results that accomplish this high spatio-temporal resolution.
    \item We combine SCI reconstruction and demosaicing for color SCI systems into a single end-to-end network.
    \item In addition to the widely used grayscale test sets, we also conduct simulation on large-scale color datasets. Furthermore, we verify the proposed network on the real data (captured by SCI cameras). Only our model can recover large scale and high compression rate SCI measurements compared with other DNN based methods thanks to the memory-efficient structure.
\end{itemize}

The rest of this paper is organized as follows. Sec.~\ref{Sec:related work} briefly reviews the related work. Sec.~\ref{Sec:videoSCI} presents the mathematical model of video SCI. Sec.~\ref{Sec:model} details our proposed model for grayscale and color video SCI reconstruction. Sec.~\ref{Sec:experiments} presents extensive results including simulation and real data.
Sec.~\ref{Sec:conclusion} concludes the entire paper.

\section{Related Work \label{Sec:related work}}

\paragraph{Video Snapshot Compressive Imaging}
Many different SCI hardware systems have been developed, by modulating the light in different approaches, \eg, usually a digital micromirror device (DMD)~\cite{Hitomi11ICCV,Ma2021_LeSTI_OE,Qiao2020_APLP,reddy2011p2c2,Sun17OE,Qiao2020_CACTI,Qiao2021_MicroCACTI} or a physical mask~\cite{Patrick13OE,Yuan14CVPR}.
Although hardware systems are mature in the laboratory, existing reconstruction algorithms are still far from
real applications.
Optimization-based methods, \eg, GAP-TV~\cite{Yuan16ICIP_GAP}, GMM~\cite{Yang14GMMonline,Yang14GMM}, DeSCI~\cite{Liu18TPAMI}, and PnP-FFDNet~\cite{Yuan20PnPSCI} consume high computational cost leading to long time reconstruction. Recently, some researchers have attempted to use deep learning in computational imaging~\cite{Iliadis18DSPvideoCS,Kulkarni2016CVPR,Meng_GAPnet_arxiv2020,Miao19ICCV,Qiao2020_APLP,2016arXivVideoCS,Yuan18OE}.
Various networks have been proposed for SCI reconstruction, and significantly reduced the running time.
However, these networks usually need a huge memory and long time for training. For instance, state-of-the-art method BIRNAT~\cite{Cheng20ECCV_BIRNAT} requires more than 32GB GPU memory (batch size is 3 and costs weeks for training) to train the model of size $256 \times 256 \times 8$.
Such a memory unfriendly model is not satisfying the increasing resolution in daily life, where HD and UHD videos are becoming widely used.
Different from previous methods, in this work, we develop a 3D CNN based network and introduce the reversible structure to reduce the training memory without loss of performance.

\paragraph{Reversible neural network}
Flow-based generative models, \eg, NICE~\cite{dinh2014nice}, real NVP~\cite{dinh2016density}, and Glow~\cite{kingma2018glow} can jointly perform generation and inference using a shared stacked reversible structure. This means that the generative process can be easily inverted, and the inference process can be computed by the inverse of the generation function.
Specifically, for $l$-{th} blocks, given an input $h^l$, divided it into two-parts $h^l_1$, $h^l_2$, NICE~\cite{dinh2014nice} performs the simple additive affine transformations:
\begin{equation}
         h^{l+1}_1=h^l_1,\quad
        h^{l+1}_2=h^l_2+m(h^l_1),
\end{equation}
where $m(~)$ is an arbitrary function. The output is the concatenation of $h^{l+1}_1$ and  $h^{l+1}_2$. The inverse transformation can be easily computed by
\begin{equation}
         h^l_2=h^{l+1}_2-m(h^{l+1}_1), \quad
         h^l_1=h^{l+1}_1.
\end{equation}
Inspired by this simple and effective setting, Rev-Net~\cite{gomez2017reversible} introduces this idea into Res-Net~\cite{resnet}, which has similar performance with Res-Net in the classification task and each block includes several reversible layers. The main strength of the reversible network is that training such networks {\em does not need to save the middle activation produced by each layer}, which occupies most of the memory. During  back-propagation, the previous layer activation can be easily computed by the reversible transformation to calculate the gradient. Therefore, saving the last activation of the stacked reversible layers allows learning the parameters, which makes the memory cost reduce from $O(L)$ to $O(1)$ ($L$ is the number of the layer). {A memory-efficient learning procedure~\cite{Kellman2020Memory} inspired by the reversible networks was proposed for unfolding networks, which is easy to act on the unfolding network to reduce the training memory without loss of accuracy.} Most recently, researchers~\cite{teshima2020coupling} have proved that flow models based on affine coupling can be universal distributional approximations.

One of the bottlenecks for the SCI reconstruction network applied in the large-scale scene is the huge GPU memory consumption as mentioned before,
because the squarely increasing pixel numbers (for a larger size) make it impossible for high-resolution scenes.
Inspired by the Rev-net~\cite{gomez2017reversible}, we propose a reversible 3D CNN for large-scale video SCI reconstruction.
Specifically, we extend the original two branches additive affine transformations into {\em multi-group transformations}.
The 3D CNN will also capture the spatio-temporal correlations in the desired video, and the reconstruction results will be more consistent in different frames.

\paragraph{Demosaicing}
For color imaging, common devices usually first capture pixels by a color filter (one pixel only sampling one color energy such as red, green or blue) and then impose an interpolation algorithm to achieve a color (usually RGB) image. This process is called demosaicing. Recently, some researchs ~\cite{chen2018learning,kokkinos2018deep,Liu_2020_CVPR} developed an end-to-end network to directly obtain a color image from the raw captured image. Motivated  by this, we extend the proposed RevSCI-net to joint demosaicing and reconstruction for the color SCI system.
To our best knowledge, this is the first attempt to use a unified end-to-end deep model to directly restore an RGB video from a compressive measurement in SCI.

\begin{figure*}
		\centering
        \includegraphics[width=1.8\columnwidth]{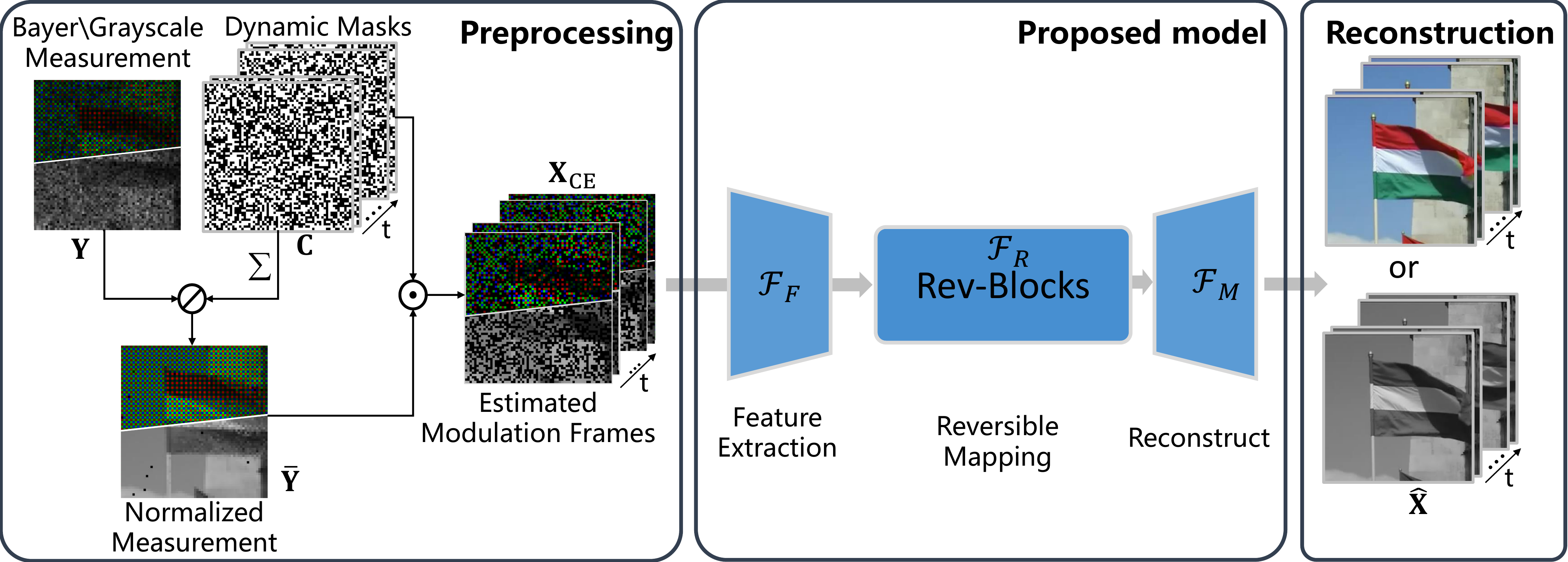}
		\caption{\label{fig:model}Proposed reconstruction pipeline. Left: the preprocessing stage to obtain the estimate of the modulated frames as the network input, which includes the information of coding masks and the normalized measurement. Middle: the RevSCI-net which concludes three parts, feature extraction, reversible non-linear mapping, and reconstruction. Right: the reconstruction video.}
\end{figure*}

\section{Video Snapshot Compressive Imaging\label{Sec:videoSCI}}
In video SCI, a dynamic scene consisting of $B$ high-speed two-dimensional frames $\{\Xmat_k\}_{k=1}^B \in {\mathbb R}^{n_x \times n_y}$
are modulated by the coding patterns (masks) $\{\Cmat_k\}_{k=1}^B \in {\mathbb R}^{n_x \times n_y}$,  respectively. These coded frames are then integrated over time on a camera, forming a {\em compressed} coded measurement (Fig.~\ref{fig:system}). The measurement $\Ymat \in {\mathbb R}^{n_x \times n_y}$ is given by
	\begin{equation} \label{Eq:YXC}
	{\textstyle \Ymat = \sum_{k=1}^B \Xmat_k \odot  \Cmat_k + \Gmat}\,,
	\end{equation}
where $\odot$ denotes the Hadamard (element-wise) product and $\Gmat \in {\mathbb R}^{n_x \times n_y}$ represents the noise. From a pixel perspective, any $B$ pixel (in the $B$ frames) at position $(i,j)$, $i = 1,\dots, n_x$; $j = 1,\dots, n_y$ are collapsed to form one pixel in the snapshot measurement by
	\begin{equation}
	{\textstyle y_{i,j} = \sum_{k=1}^B c_{i,j,k} x_{i,j,k} + g_{i,j}}\,.
	\end{equation}
Define $\xv = \left[\xv_1\tsp,\dots,\xv_B\tsp\right]$, where $\xv_k = {\rm vec}(\Xmat_k)$;
let $\Dmat_k = {\rm diag}({\rm vec}(\Cmat_k))$, for $ k=1,\dots, B$, where ${\rm vec}(~)$ vectorizes the matrix inside $(~)$ by stacking the columns and ${\rm diag}(~)$ diagonalizes the ensued vector into a diagonal matrix.
The video SCI sensing process can be written as
	\begin{equation} \label{Eq:yPhix}
	{\textstyle\yv = \Phimat \xv + \gv}\,,
	\end{equation}
where $\Phimat \in {\mathbb R}^{n\times nB}$ is the sensing matrix with  $n = n_xn_y$, $\xv\in {\mathbb R}^{nB}$ is the desired signal, and $\gv\in {\mathbb R}^{n}$ again denotes the vectorized noise.
Different from single-pixel imaging~\cite{Duarte08SPM}, the sensing matrix $\Phimat$ in~\eqref{Eq:yPhix} has a very special structure and can be written as
    \begin{equation} \label{Eq:Hmat_strucutre}
	{\textstyle\Phimat = \left[\Dmat_1,\dots, \Dmat_B\right]}\,,
    \end{equation}
where $\{\Dmat_k\}_{k=1}^B \in {\mathbb R}^{n \times n}$ are diagonal matrices of masks.
Therefore, the compressive sampling rate in  SCI is equal to  $1/B$.
Recently, researchers~\cite{Jalali19TIT_SCI} proved that high quality reconstruction is achievable when $B>1$.

In terms of color video SCI system, we consider the Bayer pattern filter sensor, where each pixel only captures the red (R), green (G) or blue (B) channel in a spatial layout such as `RGGB'. Note that two green channels are used due to the sensitivity of the human eyes.
In this case, $\Xmat_k$ is a mosaic frame and since the neighbouring pixels are sampling different color components, the values are not necessarily continuous.
To cope with this issue, previous studies~\cite{Liu18TPAMI,Yuan20PnPSCI,Yuan14CVPR} usually divide the original measurement $\Ymat$ into four-channel sub-measurements corresponding to the Bayer-filter $\{\Ymat^{r}, \Ymat^{ g1}, \Ymat^{ g2}, \Ymat^{ b}\}\in {\mathbb R}^{\frac{n_x}{2}\times \frac{n_y}{2}}$ for the R, G1, G2 and B components.
Similarly, the mask and desired signal are also divided into four components. They reconstruct each sub-signal separately using the corresponding measurement and mask and then perform demosaicing (using off-the-shelf tools) in the recovered sub-videos to generate the final color (RGB) video.

\section{The Proposed Model \label{Sec:model}}
Given the compressed measurement $\Ymat$ and coding pattern $\{\Cmat_k\}_{k=1}^B$ captured by the SCI system, the goal of the proposed model RevSCI-net is to predict the desired high-speed frames $\{\Xmat_k\}_{k=1}^B$, in other words, to learn a mapping from $\Ymat$ to $\{\Xmat_k\}_{k=1}^B$.
In this section, the details of the model will be described. Overall, our proposed model consists of three parts as shown in the middle of Fig.~\ref{fig:model}: 1) Feature extraction ${\cal F}_F$ uses several 3D CNN layers to capture the high-dimensional features of the input.
2) Feature level nonlinear mapping employs several reversible blocks ${\cal F}_R$ to transform the input features into the desired reconstruction domain.
3) Reconstruction ${\cal F}_M$ integrates the features to reconstruct the final video.

\subsection{Model for Grayscale SCI system}
\subsubsection{Feature Extraction}
Considering the measurement $\Ymat$ being a 2D matrix, we first normalize the original measurement and then combine masks and the normalized measurement to produce coarse estimates of modulated frames as follows:
\begin{equation} \label{Eq:input}
	\textstyle \overline{\Ymat} =   \Ymat \oslash \sum_{k=1}^B \Cmat_k, \quad  \Xmat_{CE}= \overline{\Ymat} \odot \Cmat,
\end{equation}
where $\oslash$ denotes the matrix dot (element-wise) division, and coarse estimates $\Xmat_{CE} \in {\mathbb R}^{B\times  n_x\times n_y }$.

After obtaining $\Xmat_{CE}$, we employ four 3D convolutional layers expressed as $\mathcal{F}_{F}$ (the kernel size is of $5\times5\times5,3\times3\times3,1\times1\times1$, and $3\times3\times3$) to extract the feature as:
\begin{equation} \label{Eq:feature extractation}
	\Hmat_{f}=\mathcal{F}_{F}(\Xmat_{CE}),
\end{equation}
where $\Hmat_{f} \in {\mathbb R}^{c_1 \times B\times  n_x\times n_y }$ is a 4D tensor and $c_1$ is the channel number.
Here, we set the stride of the final layer to 2, which reduces the resolution of the feature map by half to reduce the computational complexity.
We apply the LeakyReLU~\cite{Maas2013RectifierNI} on each convolutional layer, and do not use the batch normalization following previous research on image deburring~\cite{Nah_2019_CVPR,Lim_2017_CVPR_Workshops} and video SCI~\cite{Cheng20ECCV_BIRNAT}.
After the feature extraction operation, we obtain the coarse features of the input modulated frames.

\begin{figure*}
		\centering
        \includegraphics[width=1.5\columnwidth]{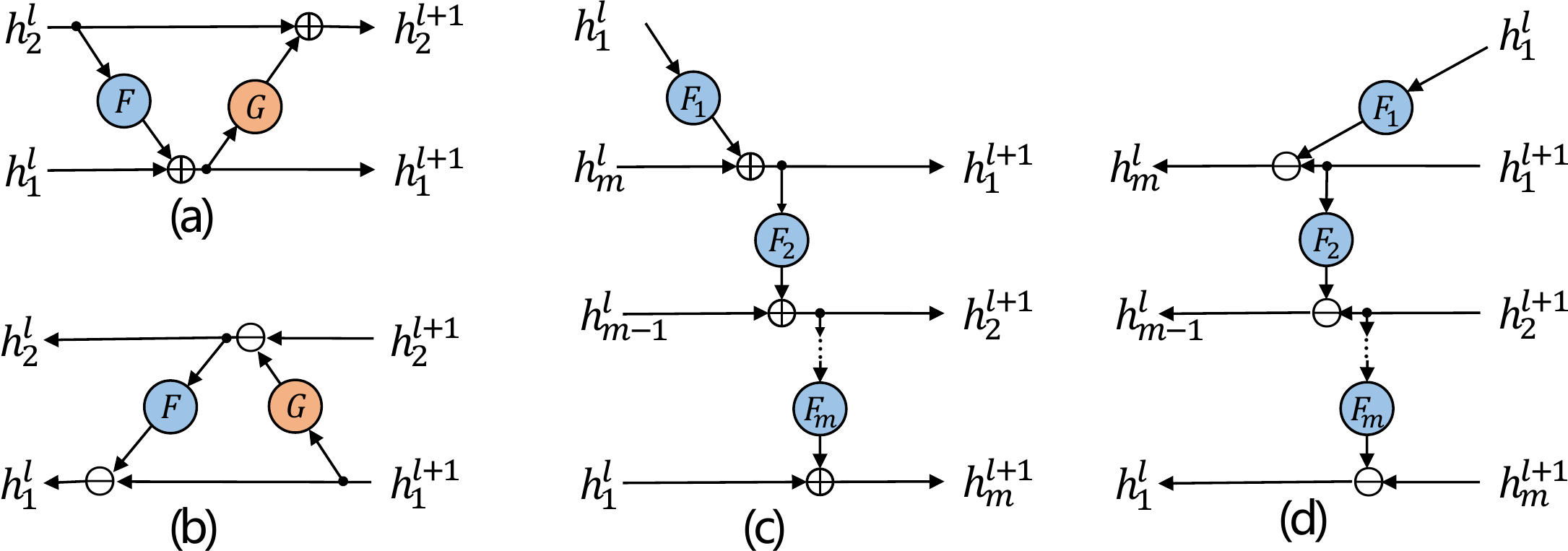}
		\caption{\label{fig:reversible block} (a) and (b) are the forward and the reverse computations of the original reversible layer in Rev-Net~\cite{gomez2017reversible}, respectively. (c) and (d) are the forward and the reverse process of the proposed multi-group reversible block, respectively.}
\end{figure*}

\subsubsection{Reversible Non-linear Mapping}
Having obtained the features of the input, we use stacked reversible blocks to transform them to the video domain features. The original reversible block in Rev-Net~\cite{gomez2017reversible} splits the input features into two parts by channel, and the transformation is:
\begin{equation} \label{Eq:reversible func}
        \hv^{l+1}_1=\hv^l_1 +\mathcal{F}(\hv^l_2),\quad
         \hv^{l+1}_2=\hv^l_2+\mathcal{G}(\hv^{l+1}_1),
\end{equation}
where $\hv^l_1, \hv^l_2,\hv^{l+1}_1, \hv^{l+1}_2\in {\mathbb R}^{\frac{c_1}{2} \times B\times  n_x\times n_y }$, and $\mathcal{F}$ and $\mathcal{G}$ are arbitrary functions. Inspired by the group convolution~\cite{krizhevsky2017imagenet,xie2017aggregated}, we modify the formulation and extend it to a multi-group reversible transformation. As shown in Fig.~\ref{fig:reversible block}(c), we split the feature into multiple parts, and the forward  function is now:
\begin{equation} \label{Eq:multi-reversible func}
    \begin{aligned}
        & \hv^{l+1}_1=\hv^l_m +\mathcal{F}_{1}(\hv^l_1),\\
        & \hv^{l+1}_2=\hv^l_{m-1} +\mathcal{F}_{2}(\hv^{l+1}_1),\\
        &  \qquad \vdots \\
        & \hv^{l+1}_m=\hv^l_{1} +\mathcal{F}_{m}(\hv^{l+1}_{m-1}),\\
    \end{aligned}
\end{equation}
where $m$ is the number of groups, and $\mathcal{F}_{*}$ can be an arbitrary function. In our experiments, we set it to two 3D convolutional layers with the kernel size of $3\times3\times3$.
With the additional dimension on the group, we extend the original reversible form, and experimental results show that these changes have improved the performance.
The inverse of the multi-group reversible transformation is thus
\begin{equation} \label{Eq:multi-reversible inverse func}
    \begin{aligned}
        & \hv^{l}_1=\hv^{l+1}_m -\mathcal{F}_{m}(\hv^{l+1}_{m-1}),\\
        & \hv^{l}_2=\hv^{l+1}_{m-1} -\mathcal{F}_{m-1}(\hv^{l+1}_{m-2}),\\
        & \qquad \vdots\\
        & \hv^{l}_m=\hv^{l+1}_{1} -\mathcal{F}_{1}(\hv^{l}_1).\\
    \end{aligned}
\end{equation}
By stacking $L$ reversible blocks, the input feature will be transformed into the reconstruction domain as:
\begin{equation} \label{Eq:feature transform}
	\Hmat_{r}=\mathcal{F}_{R}(\Hmat_{f}).
\end{equation}

Note that during the back-propagation, we only save the last activation in $\mathcal{F}_{R}$, and activations of others can be computed by the~\eqref{Eq:multi-reversible inverse func} so that calculate the gradient to update the network parameters by the chain rule. For traditional convolutional layers, adding more layers is beneficial for the non-linearity and the performance, but it will significantly increase the activation memory of the model. Fortunately, due to the reversible structure, adding the number of layers will not increase the memory cost of activations in RevSCI-net.

\subsubsection{Reconstruction}
After the reversible non-linear transformation, the goal of the reconstruction stage is to integrate the features to obtain the desired video. We utilize four 3D convolutional layers (with the kernel size of $3\times3\times3,3\times3\times3,1\times1\times1$, and $3\times3\times3$) to reduce the channel to one and achieve the final reconstruction video, \ie,
\begin{equation} \label{Eq:reconstrcution}
	\widehat{\Xmat}=\mathcal{F}_{M}(\Hmat_{r}).
\end{equation}

\subsection{Model for Color SCI System}
As mentioned before, color SCI systems capture the mosaic Bayer measurement as shown in Fig.~\ref{fig:system}. Inspired by the success of deep learning demosaicing and grayscale SCI reconstruction respectively, we conduct joint demosaicing and reconstruction using the proposed model. 

To avoid mixture of different color channels, we first separate the coarse estimates of modulated frames obtained by \eqref{Eq:input} into four individual parts corresponding to the Bayer-filter, one for red, one for blue, and two for green,
\begin{equation} \label{Eq:RGB input}
	\begin{aligned}
	\Xmat^{color}_{CE} = &[\overline{\Ymat}^r\odot\Cmat^r_1\,,...,\overline{\Ymat}^r\odot  \Cmat^r_B;\\
	&~~\overline{\Ymat}^{g1}\odot\Cmat^{g1}_1\,,...,\overline{\Ymat}^{g1}\odot  \Cmat^{g1}_B;\\
	&~~\overline{\Ymat}^{g2}\odot\Cmat^{g2}_1\,,...,\overline{\Ymat}^{g2}\odot  \Cmat^{g2}_B;\\
	&~~\overline{\Ymat}^{b}\odot\Cmat^{b}_1\,,...,\overline{\Ymat}^{b}\odot  \Cmat^{b}_B
	]_3,
\end{aligned}
\end{equation}
where $\Xmat^{color}_{CE} \in {\mathbb R}^{4 \times B \times \frac{n_x}{2} \times \frac{n_y}{2}}$ includes four color channel modulation information and superscripts $r$, $g$ and $b$ denote the red, green and blue channels, respectively.


These color independent estimates $\Xmat^{color}_{CE}$, are fed into the network. Because of the differences on the channel and the spatial resolution of the input compared with the grayscale SCI, we change the number of kernels on the first convolutional layer, and set the stride to 1 on the feature extraction stage to keep the resolution unchanged.
For reconstruction, because the color image is 3 channels, we modify the number of the kernel on the last convolutional layer.
In this manner, we extend RevSCI-net to directly obtain an RGB color video from the Bayer measurement.

\subsection{Training}
\subsubsection{Loss Function}
We jointly train our proposed model with {mean square error} (MSE) loss, \ie
\begin{eqnarray} \label{Eq:loss}
\textstyle \mathcal{L}_{\rm MSE} = \textstyle \frac{1}{c \Bmat n_x n_y}\sum_{k=1}^{B}|| \widehat{\Xmat}_{k} - \Xmat_{k} ||_2^2\,,
\end{eqnarray}
where  $\widehat{\Xmat}_{k}$ is the final reconstruction from RevSCI-net,
and $\Xmat_{k}$ is the ground-truth; $c$ is the channel number of $\widehat{\Xmat}_{k}$, one for grayscale image and three for RGB image.

\subsubsection{Back-propagation}

Note that we do not directly use the automatic differentiation routine, \eg, \texttt{Loss.backward()} in PyTorch, to calculate the gradient of parameters because this will save all activations during the forward propagation and thus costs a huge memory.
Instead, for the forward pass, we directly obtain the desired reconstruction without storing the activations of reversible blocks except the last one.
As mentioned before, for back-propagation, due to the reversible block, we calculate the previous layer activation to compute the gradient of the parameters using the chain rule; for the feature extraction and reconstruction stage, we calculate the gradient as usual.
{Thereby, during training, we only save the full activations of the feature extraction stage and the reconstruction stage (each has only four layers), and the last layer of reversible blocks whatever the number of blocks.}
\begin{table*}[htbp!]
  \caption{The average results of PSNR in dB (left entry), SSIM (right entry) and running time per measurement/shot in seconds by different algorithms on six grayscale benchmark datasets. The best results are \textbf{bold}, and the second best results are \underline{underline}.}
 \vspace{-3mm}
\begin{center}
  \resizebox{1.0\textwidth}{!}{
  \begin{tabular}{ccccccccc}
    \toprule
    Algorithm & \texttt{Kobe} & \texttt{Traffic} &\texttt{Runner} &\texttt{Drop} &\texttt{Aerial} &\texttt{Vehicle} &\texttt{Average} & Time\\
    \midrule
    GAP-TV &26.45, 0.845 &20.89, 0.715  &28.81, 0.909 &34.74, 0.970 &25.05, 0.828  &24.82, 0.838 &26.79, 0.858 &4.2\\
    DeSCI  &\underline{33.25}, \underline{0.952}  &{28.72}, 0.925  &\underline{38.76}, \underline{0.969}  &\textbf{43.22},  \textbf{0.993}  &25.33,  0.860  &27.04,  0.909  &32.72, 0.935 & 6180\\
    PnP-FFDNet & 30.50, 0.926 & 24.18, 0.828 & 32.15, 0.933 & 40.70, 0.989 & 25.27, 0.829 & 25.42, 0.849 & 29.70, 0.892 & 3.0\\
    E2E-CNN & 29.02, 0.861 & 23.45, 0.838 & 34.43, 0.958 & 36.77, 0.974 & 27.52, 0.882 & 26.40, 0.886 & 29.26, 0.900 & 0.023\\
    BIRNAT &32.71, 0.950 & \underline{29.33}, \underline{0.942} &38.70, 0.976 &42.28, 0.992 &\underline{28.99}, \underline{0.927} &\underline{27.84}, \underline{0.927} &\underline{33.31}, \underline{0.951} &0.16\\
    \midrule
    RevSCI-net$^8_{50}$ &\textbf{33.72},\textbf{0.957} &\textbf{30.02}, \textbf{0.949} &\textbf{39.40}, \textbf{0.977} &\underline{42.93}, \underline{0.992} &\textbf{29.35}, \textbf{0.924} &\textbf{28.12}, \textbf{0.937}  &\textbf{33.92}, \textbf{0.956} &0.19\\
    \bottomrule
  \end{tabular}}
  \end{center}
  \label{Table:Sim}
  \vspace{-3mm}
\end{table*}

\section{Experiments \label{Sec:experiments}}
In this section, we compare RevSCI-net with several state-of-the-art methods on both simulation datasets and real data captured by two different video SCI cameras.

\subsection{Data sets and Experimental Setting}
\paragraph{Training and testing datasets}
Following~\cite{Cheng20ECCV_BIRNAT}, we choose the data set \texttt{DAVIS2017}~\cite{Pont-TusetPCASG17} as the training set for all experiments.
\texttt{DAVIS2017} has 90 different scenes in total 6208 frames with two resolutions: $480 \times 894$ and $1080 \times 1920$.

To demonstrate the quantitative performance, we first evaluate RevSCI-net on six widely used grayscale simulation data sets including \texttt{Kobe, Runner, Drop, Traffic}~\cite{Liu18TPAMI}, \texttt{Aerial} and \texttt{Vehicle} \cite{Yuan20PnPSCI}. The resolution of these test sets is $256 \times 256$.
We follow the setting in~\cite{Liu18TPAMI}, eight sequential ($B=8$) frames are modulated by the shifting binary random masks $\{\Cmat_k\}_{k=1}^B$ and then collapsed into a single measurement $\Ymat$. Under this setting, we randomly crop patch cubes ($256 \times 256 \times 8$) from the original scenes in \texttt{DAVIS2017}, and obtain 26000 training data pairs with data augmentation.

In addition, we evaluate RevSCI-net on the RGB large-scale scene, \eg, \texttt{Messi}~\cite{Yuan20PnPSCI} with a resolution of $1080 \times 1920\times 3$ (here 3 denotes the RGB channels) and 24 sequential frames are modulated and integrated into a single Bayer measurement by the shifting binary random masks. We generate 2000 data pairs for training from \texttt{DAVIS2017} with the resolution of $1080\times 1920 \times 3$.

Lastly, we evaluate RevSCI-net on the measurements captured by two real SCI systems~\cite{Patrick13OE,Qiao2020_CACTI}.

\paragraph{Implementation details} We jointly train RevSCI-net on the RTX 2080Ti GPU for 100 epochs using PyTorch. Adam optimizer~\cite{kingma2014adam} is used to minimize the loss function with the starting learning rate of $2\times 10^{-4}$. Then, we reduce the learning rate by 10$\%$ every 10 epochs. It takes about a week to train the entire network. The detailed architecture for RevSCI-net is given in the supplement material (SM).

\paragraph{Counterparts and Performance Metrics} We compare RevSCI-net with five competitive counterparts: two iterative optimization methods -- GAP-TV~\cite{Yuan16ICIP_GAP} and  DeSCI~\cite{Liu18TPAMI}, and three methods using deep learning -- the plug-and-play method PnP-FFDNet~\cite{Yuan20PnPSCI} integrated the deep denoiser as a prior, E2E-CNN~\cite{Qiao2020_APLP} which is a deep CNN model, and BIRNAT~\cite{Cheng20ECCV_BIRNAT} which builds a bidirectional RNN and produces current state-of-the-art results.
For the simulation datasets, both peak-signal-to-noise ratio (PSNR) and structural similarity (SSIM)~\cite{Wang04imagequality} are used as metrics to evaluate the reconstruction quality. Besides, we give the running time at the testing stage which determines the usability of the method in real applications.

\begin{figure}[htbp!]
		\centering
        \includegraphics[width=1.0\columnwidth]{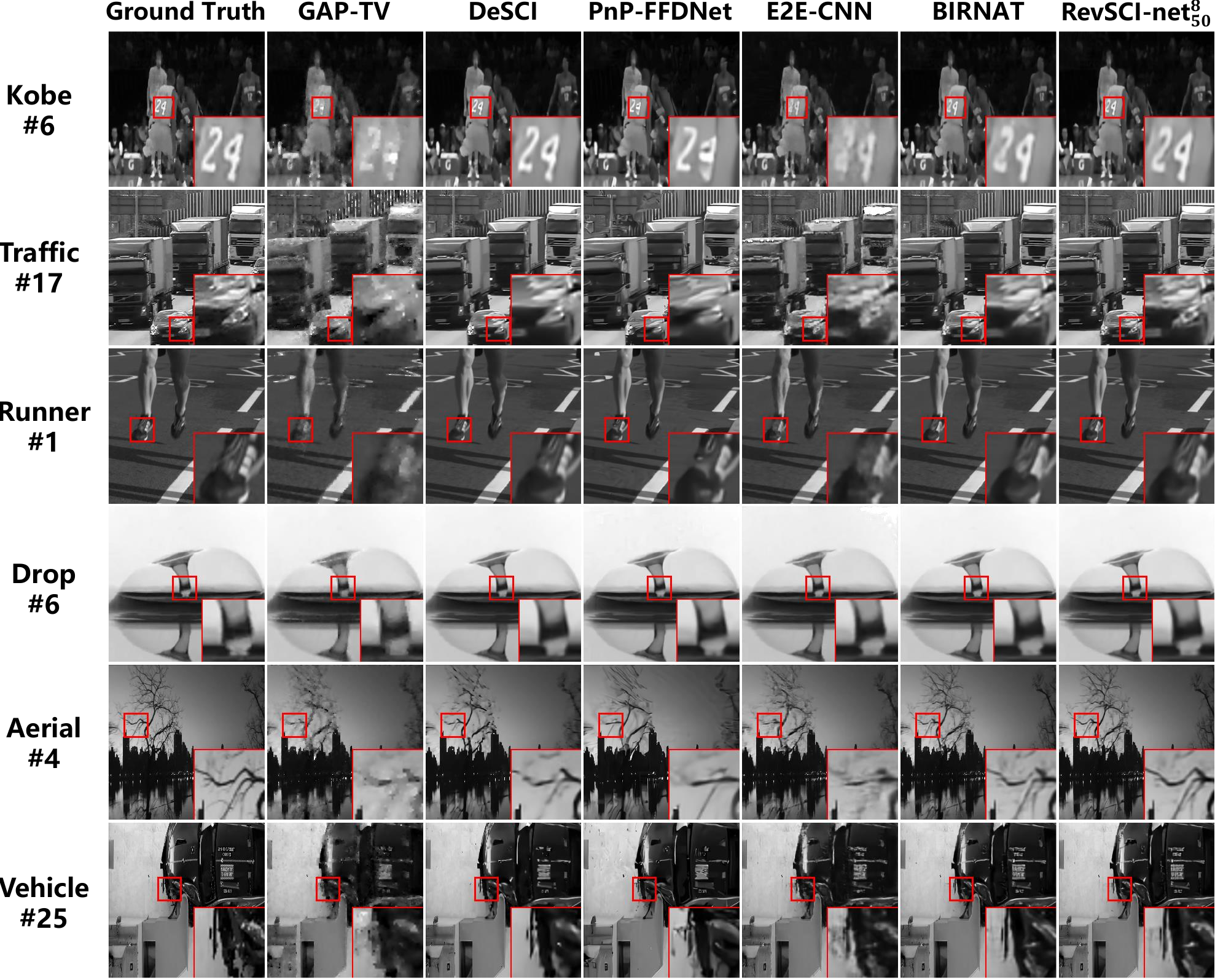}
		\caption{\label{fig:sim} Selected reconstruction frames of six grayscale benchmark datasets. Full reconstruction videos are in SM.}
\end{figure}

\subsection{Results on Simulation Datasets}
We first show the results of six grayscale datasets in Table~\ref{Table:Sim} and Fig.~\ref{fig:sim}. Table~\ref{Table:Sim} summarizes the comparisons with previous methods on PSNR, SSIM, and running time. RevSCI-net$^8_{50}$ in Table~\ref{Table:Sim} includes 50 rev-blocks and each of them are split into 8 groups. It can be observed that the proposed RevSCI-net outperforms others, specifically 0.61dB in PSNR higher than previous state-of-the-art method BIRNAT, and using a similar testing time. Fig.~\ref{fig:sim} plots the selected reconstruction frames of different methods compared with the ground truth. RevSCI-net provides cleaner and sharper reconstructions than other algorithms; the fine details are recovered accurately. Please refer to full videos in the SM.

\begin{figure}[htbp!]
		\centering
        \includegraphics[width=1\columnwidth]{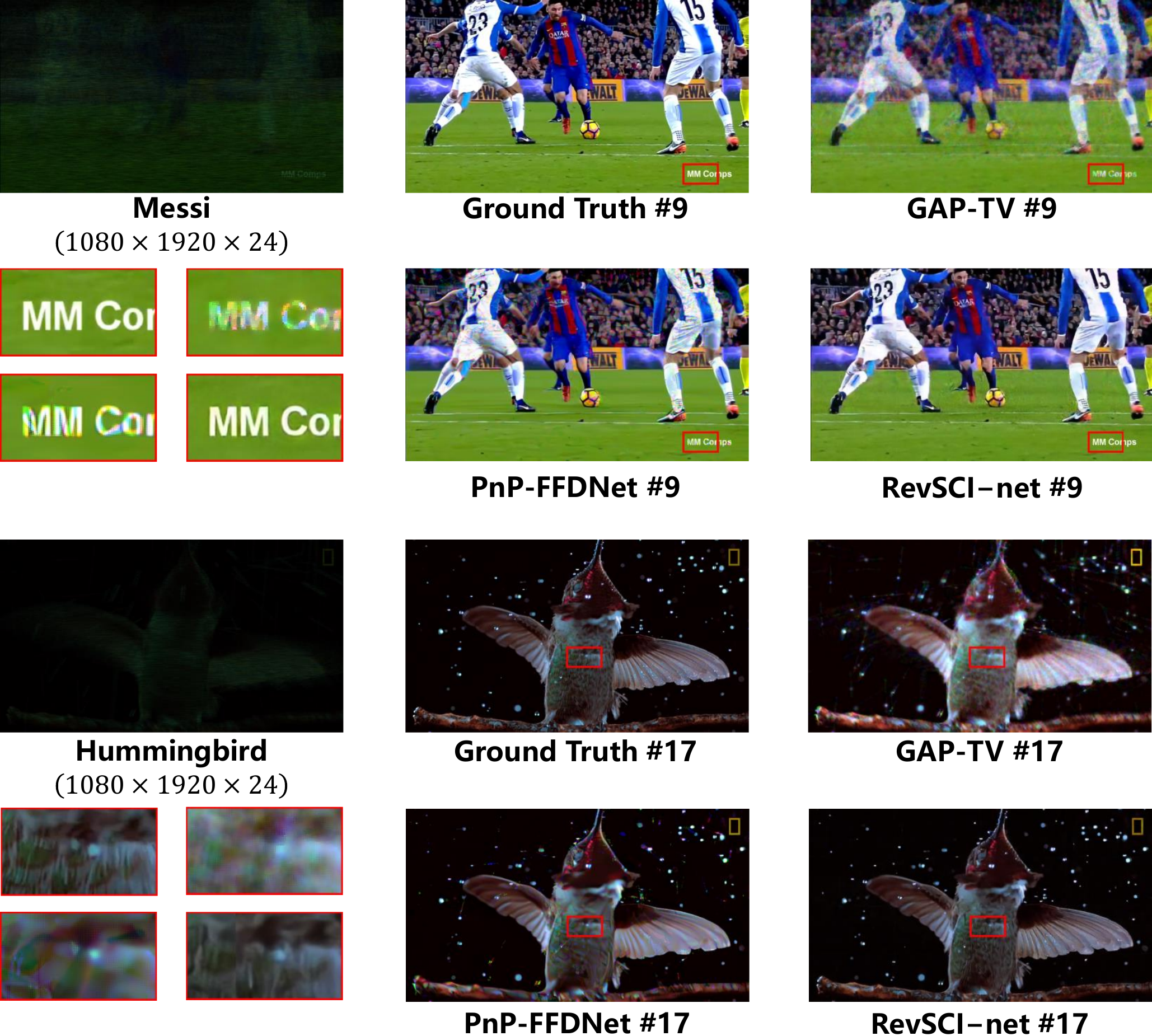}
		\caption{\label{fig:messi} The reconstruction frames of RGB large-scale dataset \texttt{Messi} and \texttt{Hummingbird}. 24 RGB frames of size $1080\times1920\times 3$ are reconstructed from a single Bayer measurement of size $1080\times 1920$.}
		\vspace{-3mm}
\end{figure}

\begin{table}[htbp!]
  \caption{ The average results of PSNR in dB (left entry), SSIM (right entry) by different algorithms on two color benchmark datasets.}
\begin{center}
\vspace{-3mm}
  \resizebox{.7\columnwidth}{!}{
  \begin{tabular}{c|c|c}
    \toprule
    Algorithm &\texttt{Messi}  &\texttt{Hummingbird}\\
    \hline
    GAP-TV &18.56, 0.7209 &18.29, 0.6449 \\\hline
    PnP-FFDNet &21.54, 0.7959 &24.13, 0.8340 \\\hline
    RevSCI-net &24.35, 0.8576 &31.97, 0.8816 \\
    \bottomrule
  \end{tabular}}
  \end{center}
  \label{Table:large}
  \vspace{-7mm}
\end{table}

Next, we show the results of RGB large-scale simulation dataset \texttt{Messi} and \texttt{Hummingbird} (1080$\times$1920$\times$3$\times$24, where $B$=24) in Fig~\ref{fig:messi} and Table~\ref{Table:large}. It worth noting that the proposed RevSCI-net is the first end-to-end training network (joint reconstruction and demosaicing) to recover such a large SCI scene. DeSCI will consume days to reconstruct, and therefore we only compare with GAP-TV and PnP-FFDNet. More analysis of memory and time is shown in Table~\ref{Table:Revmemory}. 

\begin{table}[htbp!]
\begin{center}
  \caption{\label{Table:Revmemory} Training memory occupation (MB) and running time (seconds) in videos of different resolution and compression ratio. We only show the GPU memory occupation during training on BIRNAT and RevSCI-net with a single sample. `-' means not available due to too long time or too big memory consumption.}
  \vspace{-3mm}
  \resizebox{1\columnwidth}{!}{
  \begin{threeparttable}
  \begin{tabular}{c|c|c|c|c|c}
    \toprule
    \multicolumn{2}{c|}{Method}  &256$\times$256$\times$8 & 256$\times$256$\times$14 &512$\times$512$\times$50 &1920$\times$1080$\times$24\\
    \hline
    GAP-TV & Time &4.2 &11.6 &180 &524 \\\hline
    DeSCI & Time &6180 &3185.8 &12600 &-\\\hline
    PnP-FFDNet & Time &3.0 &2.7 &88 &253\\\hline
    \multirow{2}{*}{{BIRNAT}}  &Memory &17748 &23912 &$>$48000 &$>$48000\\ \cline{2-6}
     &Time &0.16 &0.28 &- &-\\\hline
    \multirow{2}{*}{{RevSCI-net$^8_{50}$}}  &Memory &1350 &1876 &11648* &46215*\\ \cline{2-6}
     &Time &0.19 &0.33 &3.56 &12.46\\
    \bottomrule
  \end{tabular}
  \begin{tablenotes}
  \item[*] We used NVIDIA RTX8000 GPU with 48GB memory to train the model for the large-scale data.
  \end{tablenotes}
    \end{threeparttable}
    }
  \end{center}

  \vspace{-7mm}
\end{table}

\subsection{Ablation Study}
To quantitatively verify the contributions of the RevSCI-net, we modify the number of rev-blocks and groups in RevSCI-net with results shown in Table~\ref{Table:rev_layer}. The models are tested on the six grayscale datasets with results in Table~\ref{Table:Sim}. Note that stacking the rev-block will significantly increase the reconstruction quality, and adding the number of groups will help the reconstruction by more sufficiently affine transformations in the feature-level.
As mentioned before, adding rev-blocks will not increase the activation memory during training, while adding parameters will only increase a small amount of storage.

\begin{table}[htbp!]
  \caption{Computational complexity and average reconstruction quality on six grayscale test sets using RevSCI-net with different reversible blocks and groups.}
\begin{center}
\vspace{-5mm}
  \resizebox{1\columnwidth}{!}{
  \begin{tabular}{c|c|c|c|c|c}
    \toprule
    Model &Parameters ($\times10^{6}$) &MACs ($\times10^{11}$) &Memory (MB) &PSNR &SSIM\\
    \hline
    RevSCI-net$^2_{18}$ &2.11 &3.02 &1283 &33.11 &0.947\\
    \hline
    RevSCI-net$^2_{28}$  &3.22 &4.47 &1301 &33.34 &0.951\\
    \hline
    RevSCI-net$^2_{50}$  &5.65 &7.67 &1350 &33.62 &0.954\\
    \hline
    \hline
    RevSCI-net$^4_{50}$  &5.65 &7.67 &1350 &33.76 &0.955\\
    \hline
    RevSCI-net$^8_{50}$  &5.65 &7.67 &1350 &33.84 &0.956\\
    \bottomrule
  \end{tabular}}
  \end{center}
  \label{Table:rev_layer}
  \vspace{-7mm}
\end{table}

\begin{figure*}[htbp!]
		\centering
        \includegraphics[width=2\columnwidth]{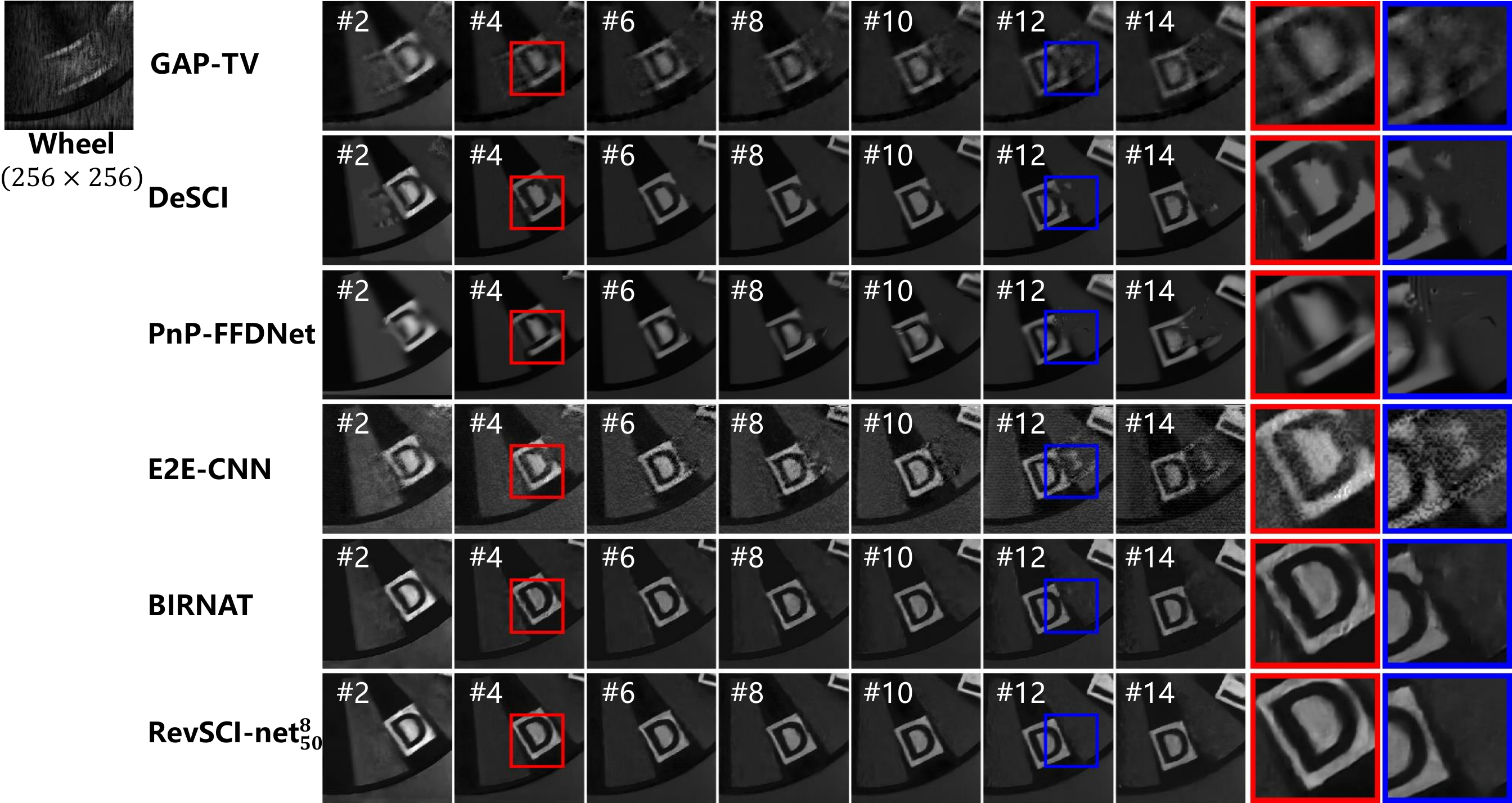}
		\caption{\label{fig:chop wheel} The reconstruction frames of real data \texttt{Wheel} with size $256\times256\times 14$. Full videos in SM.}
\end{figure*}

\begin{figure*}[htbp!]
		\centering
        \includegraphics[width=2\columnwidth]{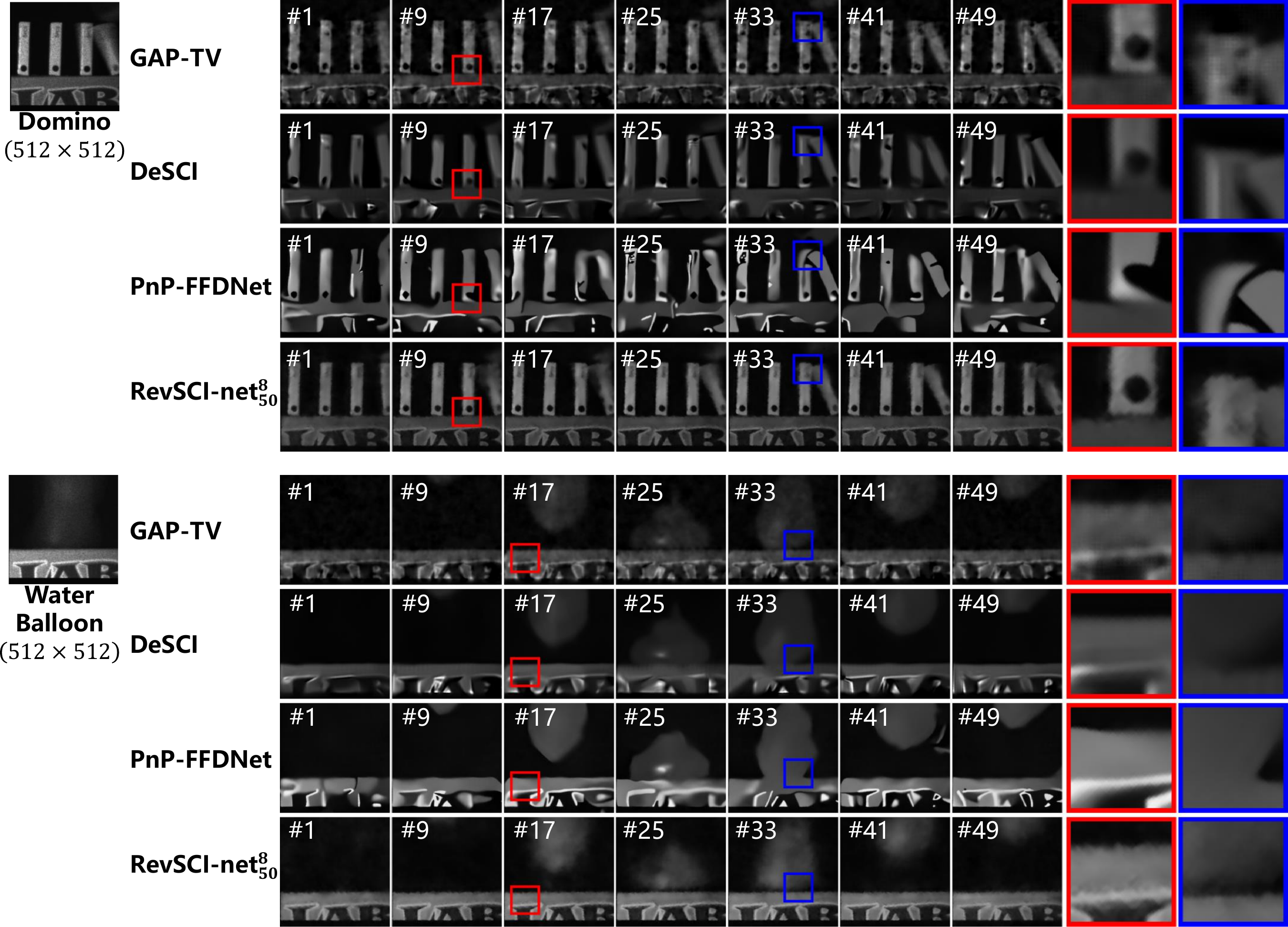}
		\caption{\label{fig:domino} The reconstruction frames of real data \texttt{Domino} and \texttt{Water Balloon} with size $512\times512\times 50$. Full videos in SM.}
\end{figure*}
\subsection{Results on Real Datasets}
We now apply the proposed RevSCI-net on real data captured by two SCI cameras~\cite{Patrick13OE,Qiao2020_CACTI}. The results of \texttt{Wheel} with a size of 256$\times$256$\times$14 are shown in Fig.~\ref{fig:chop wheel}. It can be observed that the results of RevSCI-net provide sharper edges and clearer letter `D' than others. The results of \texttt{Domino} and \texttt{Water Balloon} with a size of 512$\times$512$\times$50 are shown in Fig.~\ref{fig:domino} with full videos in the SM. In such a large compression ratio (50), the results of DeSCI are extremely over smooth, and GAP-TV introduces significant noise.
Unpleasant artifacts exist in the results of PnP-FFDNet.
The results of RevSCI-net have more accurate motions and contours.
As mentioned before, our proposed RevSCI-net is the first end-to-end deep model that can handle such a large-scale problem, while existing deep model will fail due to high demands of GPU memory.
Thanks to the reversible network, we can now apply RevSCI-net to large-scale SCI reconstruction problems in our daily life.

\section{Conclusions \label{Sec:conclusion}}
Efficient reconstruction algorithms for large scale problems have been a  long-term challenge in inverse problems.
Inspired by the recent advances of deep learning, fast inference is promising by training a deep network. However, for real life large-scale problems, deep networks are usually starving for memory and power.
In this paper, based on the application of video snapshot compressive imaging, we propose a novel memory efficient network for large-scale reconstruction. Specifically, we introduce the reversible 3D CNN in SCI reconstruction, and build the memory-efficient RevSCI-net. For the first time, we have achieved end-to-end training network to recover FHD SCI measurements. In addition, we combine demosaicing and SCI reconstruction to directly obtain RGB videos from raw Bayer measurements and thus pave the way of real applications of SCI~\cite{Lu20SEC}. Extensive results demonstrated that RevSCI-net has significant improved reconstruction quality and running time. Besides video SCI, we believe RevSCI-net will work well in other computational imaging problems such as compressive spectral imaging~\cite{Meng20ECCV_TSAnet,Meng2020_OL_SHEM,Yuan15JSTSP}.

Another way to apply CNN to large scale data is to train a small network but to adapt it to different modulation masks. One recent work has been done in~\cite{Wang2021_CVPR_MetaSCI} demonstrating the promise of this direction using meta learning. As mentioned in~\cite{Yuan2021_SPM}, the other line of work is using deep unfolding~\cite{Meng_GAPnet_arxiv2020}. The work in~\cite{Huang2021_CVPR_GSMSCI} unfolds the Gaussian scale mixture model and is able to train a small-size but multi-stage network to be used in the large scale spectral SCI problem~\cite{Meng20ECCV_TSAnet,Zheng20_PRJ_PnP-CASSI}.

\section*{Acknowledgement}
B. Chen acknowledges the support of NSFC (61771361), the 111 Project (No. B18039), and the Program for Oversea Talent by Chinese Central Government.

{\small
\bibliographystyle{ieee_fullname}

}

\end{document}